\documentclass[prb,twocolumn,showpacs,preprintnumbers,amsmath,amssymb]{revtex4}
\usepackage{graphicx}
\usepackage{dcolumn}
\usepackage{bm}

\begin{document}

\preprint{APS/Manuscript}

\title{ $O(N)$ Algorithms in TBMD Simulations of the Electronic Structure 
of Carbon Nanotubes  }

\author{ G. Dereli}
\affiliation {Department of Physics, Middle East Technical University, 06531 Ankara, Turkey}
\email {gdereli@metu.edu.tr}
\author{C. \"{O}zdo\u{g}an}
\affiliation {Department of Computer Engineering, \c{C}ankaya University, 06530 Ankara, Turkey  } 
\email{ ozdogan@cankaya.edu.tr}

\date{\today}

\begin{abstract}
The $O(N)$ and parallelization techniques 
have been successfully applied in tight-binding MD simulations
of SWNT's of various chirality.
The accuracy of the $O(N)$ description is found to be  enhanced by the use of basis functions of neighboring atoms (buffer). The importance of buffer size in evaluating the simulation time, total energy and force values together with electronic temperature has been shown.
Finally, through the local density of state results,
the metallic  and semiconducting behavior of 
$(10 \times 10)$ armchair and $(17 \times 0)$ zig zag SWNT's, respectively, 
  has been demonstrated.
\end{abstract}

\maketitle

\section{Introduction}

The carbon nanotubes are playing a major role in the design of next generation nanoelectronic
and  nanoelectromechanical devices due to their novel mechanical and electronic properties \cite{sadrdr}. The conductivity behavior of single walled carbon nanotubes (SWNT) is mostly determined by the chirality of the tubes.
Depending on its chirality, a SWNT  could be a conductor, semiconductor, 
or insulator.
It is now widely known that the conductivity of the tubes may also change 
due to the presence of defects as well as radial deformations \cite{new1}. 
Tight binding calculations have shown that deformations such as uniaxial compressive/tensile or torsional will also
modify the band gap  of the nanotubes and under such deformations SWNT undergo 
conducting-semiconducting-insulator transitions \cite{yang1, yang2}.
Real space algorithms has been successfully used to perform the {\sl ab initio} 
electronic structure calculations in the literature \cite{nardelli1}--\cite{ozaki1}.
The main objective of this paper is the use of $O(N)$ parallel tight binding molecular dynamics method in studying the electronic structure of SWNT's with diameters 
going upto 2nm. We applied the $O(N)$ and parallelization techniques in particular to 
$(10 \times 10)$ and $(17 \times 0)$ SWNT's.

The tight binding theory (TB) has been established as a good comprimise between {\sl ab initio} simulations and model-potential ones, bridging the gap
between them, either as far as the overall numerical efficiency and/or as far as the accuracy were concerned.
Tight Binding Molecular Dynamics (TBMD) is a computational tool designed to run finite-temperature MD simulations within the semi-empirical tight-binding scheme \cite{colombo1}, \cite{colombo2}.
This technique can be used to simulate material systems 
at different conditions of temperature, pressure, etc., including materials at extreme thermodynamical conditions.
The electronic structure of the simulated system can be  calculated by a TB Hamiltonian  so that the quantum mechanical many-body nature of interatomic forces is naturally taken into account.
The traditional TB solves the Schr\"{o}dinger equation by direct matrix diagonalization, which results in cubic scaling with respect to the number of atoms ( $O(N^3)$).
The $O(N)$ methods, on the other hand,  make the approximation that only the local environment contributes to the bonding and hence band energy of each atom.
In this case the run time would be  in linear scaling with respect to the number of atoms. Moreover, $O(N)$ schemes can be efficiently parallelized through the use of message passing libraries. 
The message passing libraries such as PVM and MPI  
are making the simulations possible on clusters of computers.
We applied these two techniques ($O(N)$ and parallelization)
 succesfully to the SWNT simulations on a cluster of eight PC's.
Details of the computational study can be found in our previous work
\cite{ozdeca},\cite{ozdogan}.   

\section{The Method}

Traditional TB solves the Schr\"{o}dinger equation in the reciprocal space
by direct matrix diagonalization, which results in cubic scaling with
respect to the number of atoms. The O(N) methods on the other hand, solve for the band
energy in real space and make the approximation that only the local
environment contributes to the bonding, and hence band energy, of
each atom. All the O(N) methods in which the properties of  the whole system are
computed (such as the charge distribution, the total energy or the forces on
all atoms), provide necessarily approximations to the exact solution of the effective
one--electron Hamiltonian. These approximations are based on physical assumptions,
which are generally connected to the above mentioned locality or nearsightedness
principle in one way or another. Most of the implementations of the
O(N) procedure have been developed for the orthogonal tight--binding
Hamiltonian. The O(N) techniques may be roughly grouped into two
categories: variational methods and moment--based methods. There are two
types of variational methods: the density matrix methods and
localized orbital methods. There is also a variety of moment methods. The
O(N) scaling, in these approaches,
arises from the decay and/or truncation  of these respective
quantities  \cite{ordejon}- \cite{boaogo}. 
In our work, we adopted a divide and conquer approach (DAC) first proposed by Yang
\cite{yang}-\cite{zhpaya} as a linear--scaling method used to carry
out quantum calculations. The basic strategy of this method is as follows: divide
a large system into many subsystems, determine the electron density
of each subsystem separately, and sum the corresponding contributions
from all subsystems to obtain solely from the electron density \cite{leleya}.
Each subsystem is described by a set of local basis functions, instead of the entire
set of atomic orbitals. The accuracy of the description is enhanced by the use of
basis functions of neighboring atoms. These neighboring atoms are called the
{\it{buffer\/}}.  Here the form of the  
Schr\"{o}dinger's equation of the buffer will be 
as in Ref. \cite{ozdeca}. The eigenvalues and vectors are found by diagonalizing the Hamiltonian matrices for  each subsystem. Let
${\cal{N}\/}$ be the number of atoms in the buffer region while
$N$  the number of atoms in a subsystem and 
$NCell$ the number of subsystems. A subsystem will be labeled by
$\alpha$. 
$P^i$ shows the projection of the i'th electron and 
$O^i\equiv f(({\cal E}_i-\mu)/k_BT)$ the occupation.
$n$ will be the total number of atoms in the system,
${\cal{NN}}$ the number of atoms in the buffer region
that are in the interaction distance (cutoff).
Thus we write 
\begin{equation}
P^i=\sum_{j=1}^{4N} |H(j,i)]^2
\end{equation}
where $H(j,i)$ is the  $ji^{th}$ eigenvector after diagonalization scheme and
\begin{equation}
O^i={2 \over 1+ f(({\cal E}_i-\mu)/k_BT)} .
\end{equation}
Then
\begin{equation}
\rho^i_\alpha \equiv P^i* O^i= {2 \over 1+f(({\cal E}_i-\mu)/k_BT)}
*\sum_{j=1}^{4N} |H(j,i)]^2
\end{equation}
and the  subsystem density
\begin{equation}
\rho_\alpha=\sum^{4\cal{N}\/}_{i=1} \rho^i_\alpha .
\end{equation}
The trace
\begin{equation}
trace=\sum^{NCell}    \rho_\alpha
\end{equation}
must be equal to number of electrons in the system so that the error 
\begin{equation}
error=trace-4*n.
\end{equation}
In the above expressions, 
$f(x)=1/(1+exp(x))$ is the Fermi function, $\mu$ is the chemical potential for the electrons,
$k_B$ is the Boltzmann constant, and $T$ is the temperature of the electrons
in  degrees Kelvin.
In case  the error exceeds the accuracy needed within the desired electron
temperature, the chemical potential is recalculated from
\begin{equation}
\mu_{new}={-error \over d \rho}+\mu
\end{equation}
where
\begin{equation}
d \rho=\sum^{NCell} \sum_{i=1}^{4\cal{N}\/}
\left[  (O^i*P^i)*(1-O^i)/k_BT\right] .
\end{equation}
This procedure is repeated until the desired level of accuracy is reached. The final
value for the chemical potential is identified as the Fermi energy level of the system.

The band structure energy of the system is calculated as
\begin{equation}
E_{bs}=\sum^{NCell}_{\alpha=1} ebstot_{\alpha}
\end{equation}
where  $ebstot_{\alpha}$ shows the contribution of a subsystem to the band
structure energy of the system:
$$
ebstot_{\alpha}
=\sum_{j=1}^{4N} [ \left(\sum_{j=i+1}^{4\cal{N}\/}
 2*density_{\alpha}(i,j)*{\cal{H}\/}(i,j)\right)
$$
\begin{equation}
+ density_{\alpha}(i,i)*{\cal{H}\/}(i,i)]
\end{equation}
where
$$
density_{\alpha}(k,j)=\sum_{j=1}^{4N}[\left(\sum_{k=j}^{4N} \sum_{i=1}^{4\cal{N}\/}
H(j,i)*H(k,i)*O^i\right)+
$$
\begin{equation}
\left(\sum_{k=4N+1}^{4\cal{N}\/} \sum_{i=1}^{4\cal{N}\/}
0.5*H(j,i)*H(k,i)*O^i\right)]
\end{equation}
and ${\cal{H}\/}(i,j)$ is the  $ji^{th}$ element of the Hamiltonian matrix of the subsystem.

The next step is to find the forces that each atom experiences arising from the
electronic structure, i.e. in the x-direction:
\begin{equation}
f_{x_{j=1\ldots n}}= \sum^{NCell}_{\alpha=1} \sum_{i=1}^{\cal{N}\/} f^{\alpha}_{x_i}
\end{equation}
where
\begin{eqnarray}
f^{\alpha}_{x_i}=&&\sum_{j=1}^{\cal{NN\/}}\sum_{im=1}^4\sum_{jm=1}^4
density_{\alpha}(4(i-1)\nonumber\\
&&+im,4(j-1)+jm)*Force(im,jm)
\end{eqnarray}
and $Force(im,jm)$ has the same form
as in Ref. \cite{ozdeca}. 
Total energy of the system has the form,
\begin{equation}
E_{tot}=E_{bs}+E_{rep}
\end{equation}
where $E_{rep}$ has the same form
also as in Ref. \cite{ozdeca}. 
The energetics and forces are now calculated and then
molecular dynamics scheme is applied and
this procedure is continued until structural
stability is sustained.

\section{Results and Discussion}

We applied the $O(N)$ and parallelization techniques in particular to 
$(10 \times 10)$ and $(17 \times 0)$ SWNT's.
Obviously the results obtained with $O(N)$ algorithm must be consistent with $O(N^3)$ results for the same SWNT. Buffer size and the electronic temperature are the important $O(N)$  parameters that affects the results. We produced
the  Fermi-Dirac distribution, local density of states and energetics for the
above types of  SWNT's.   
Through these, it is possible to distinguish between the metallic and semi-conducting behavior of SWNT's depending on their chiralities.

The buffer size is an important parameter in the $O(N)$ simulations.
 Each subsystem has its own buffer region so that its own small sized Hamiltonian matrix. After diagonalizing this Hamiltonian matrix, the eigenvalues and eigenvectors are obtained. The next step is to obtain all these informations for all subsystems and then calculate the overall system property; chemical potential. This parameter gives us the value for Fermi energy level. Then, the forces that each atom experiences and the contribution of each subsystem to band structure energy of the system are calculated. All these procedures are repeated through each  MD time step.
The results obtained with O(N) algorithm must be consistent with O($N^3$) results for the same system. To ensure this, the value for the buffer size parameter must be 
investigated.

\begin{figure}
\vskip 18.0 cm
    \includegraphics{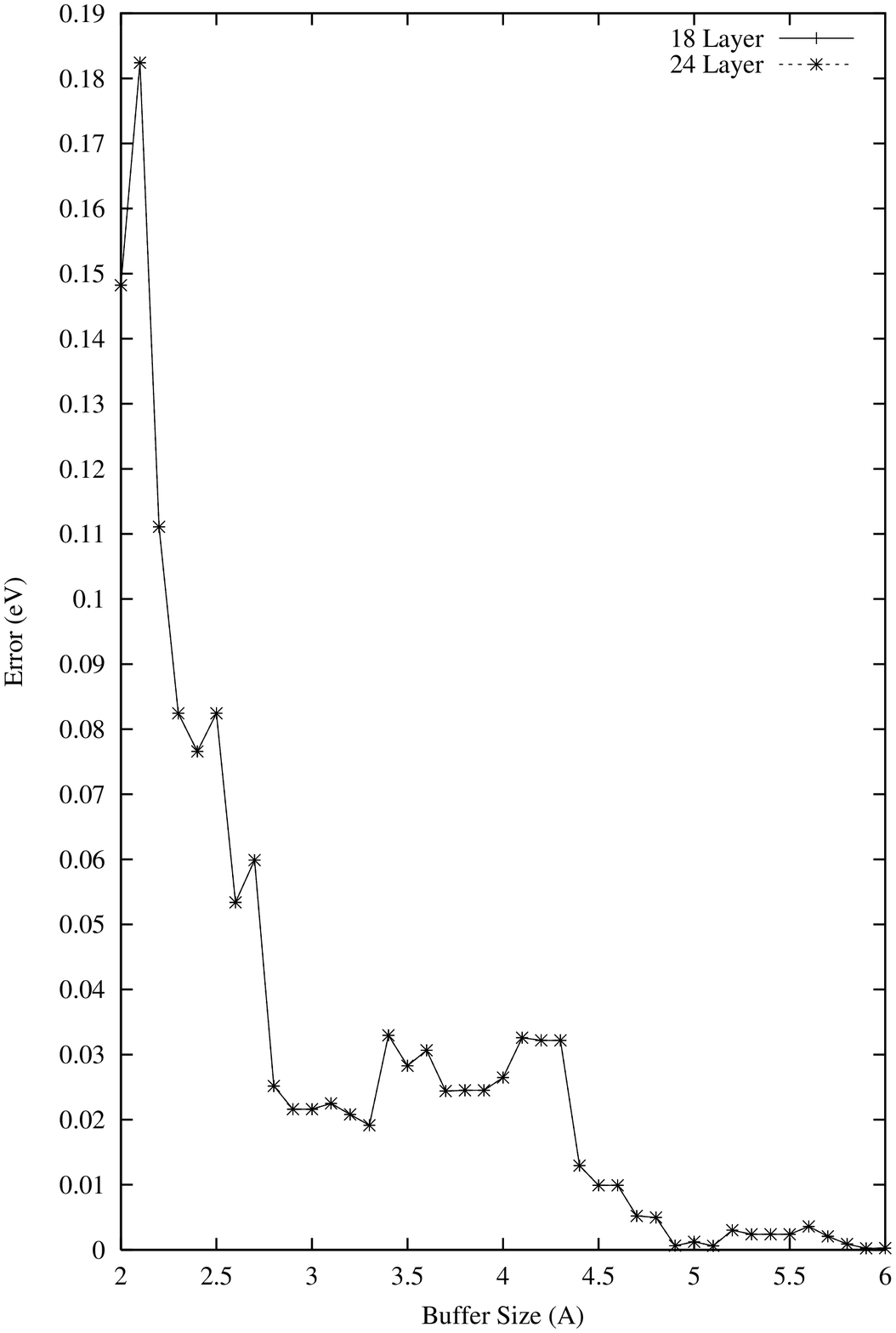}
\vspace{1.5 cm}
\caption{The difference of $O(N^3)$ total energy result (-8.350497775) with $O(N)$ total energy result  for the variation of $O(N)$ parameter ( Buffer Size) for the 24 Layers and 18 Layers 10x10 Tube with box size is equal to 1.229 \AA ~, the electronic temperature is $k_BT=0.005~eV$.}
\label{4.1.1}
\end{figure}

Another important parameter in the simulation is the cuboidal box size. We took the cube size equal to the distance between the layers in the tube so that each cube has equal amount of atoms. This also provides the same number of interacting neighbor atoms (buffer) for each subsystem. The PBC is applied in the z--direction only. Hence, the system behaves   as infinitely long tube. In  Fig. \ref{4.1.1}, it is seen that the difference with O($N^3$) total energy result for 18 layers and 24 layers are exactly same, since PBC works well. We have chosen 20 layers for both $(10 \times 10)$ and $(17 \times 0)$ tube structures for the rest of our study.

Buffer atoms are selected using a distance criterion, $R_b$. That is, if an
atom is within a distance  $R_b$ of a subsystem, this atom will be included as 
buffer atom for that subsystem. The diagonalization for a subsystem is performed with
atomic basis functions  of the subsystem atoms and buffer atoms, and the
computational effort scales as $N_\alpha^3$, where $N_\alpha$ is the number of basis
functions 
in the $\alpha$ subsystem and its buffer region. After diagonalization, the resulting eigenvalues and eigenvectors give us necessary information for local Density of States (lDOS) and for force expressions to evaluate the next MD iteration.

\begin{figure}
 \vskip 18.0 cm
    \includegraphics{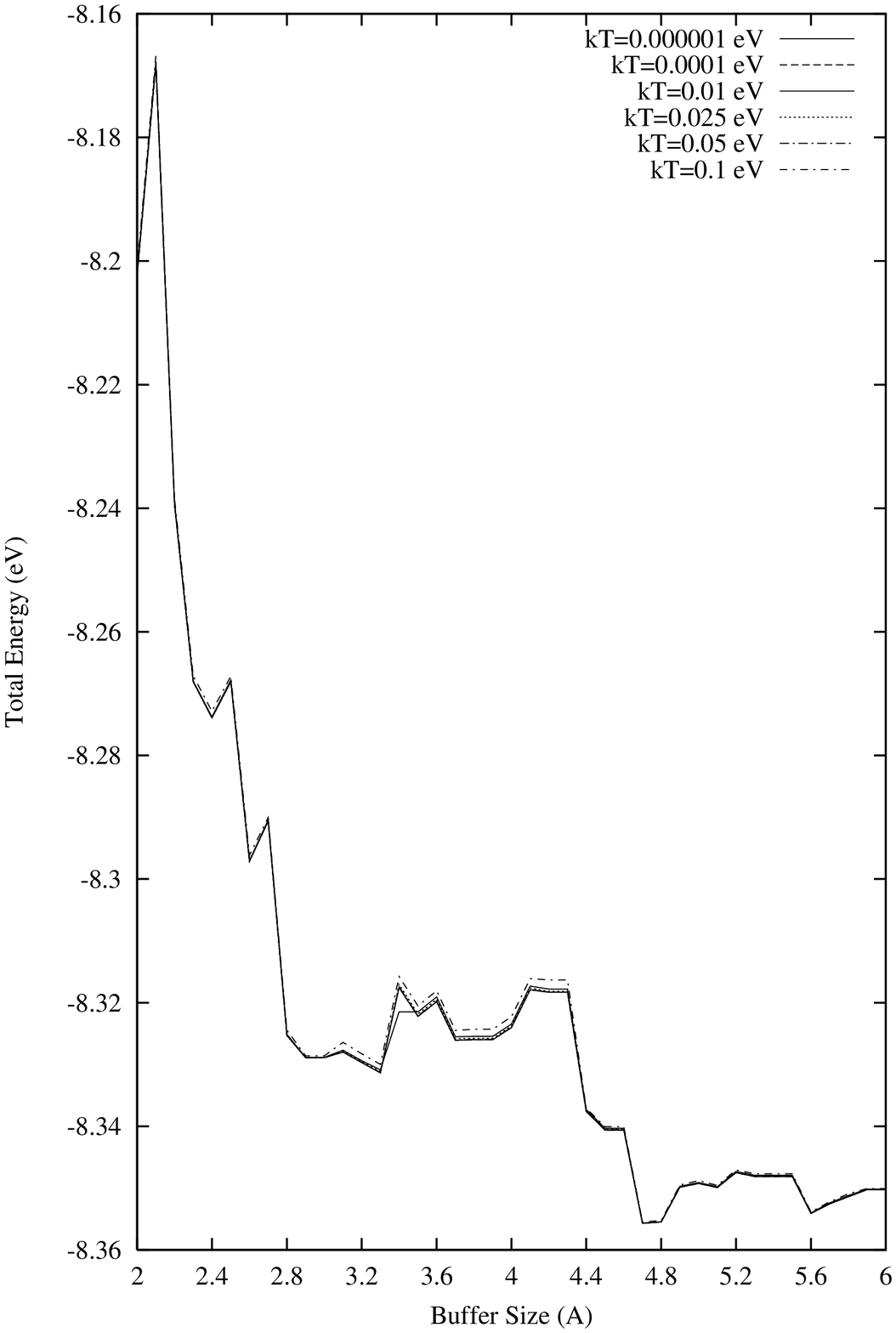}
\vspace{1.5 cm}
\caption{The effect of Buffer Size for the Total Energy
value for the tube structure 10x10,
with different electronic temperature  values,
respectively.}
\label{4.1.3}
\end{figure}

\begin{figure}
 \vskip 18.0 cm
    \includegraphics{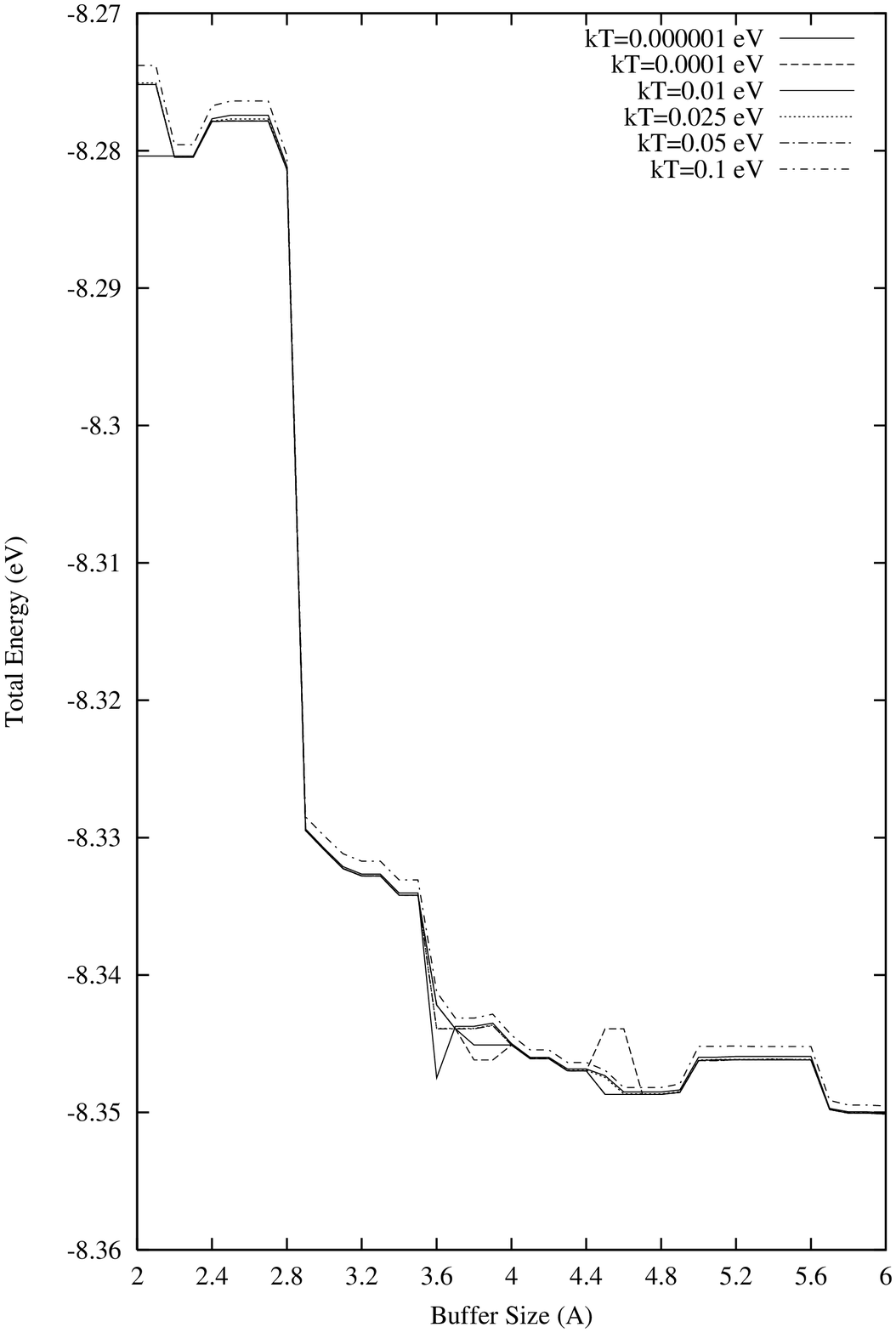}
\vspace{1.5 cm}
\caption{The effect of Buffer Size for the Total Energy
value for the tube structure 17x0,
with different electronic temperature  values,
respectively.}
\label{4.1.4}
\end{figure}

In Fig. \ref{4.1.1}, the effect of the buffer size on the total energy within the given constraints such as boxsize, electronic temperature is given. It is seen that the effect of the buffer size on the O(N) TBMD is very important. For the 10x10 SWNT
difference with O($N^3$) TBMD result (error) decreases when the buffer size is increased; then reaches to desired accuracy and fluctuates around this value. Buffer size  is important in evaluating the simulation time, energy and force values. Such as, if the buffer size is chosen a higher value than necessary, it will affect the simulation time in cubic manner since the Hamiltonian matrix is constructed with respect to the number of interacting atoms in the buffer region. On the other hand, if this parameter is chosen a low value then it will not be able to to produce the correct energy and force values. In Figs. \ref{4.1.3} and \ref{4.1.4}, the effect of the buffer size on the O(N) TBMD for the $(10 \times 10)$ and $(17 \times 0)$ tube structures together with the effect of the electronic temperature are also given. From Fig. \ref{4.1.3} the buffer size for $(10 \times 10)$  tube is chosen as 4.9 \AA~ and for $(17 \times 0)$ tube it is
5.7 \AA~ ( see Figure \ref{4.1.4}).
 It is important to keep the buffer size parameter as small as possible and at the same time, it must be able to produce the same values with the O($N^3$) TBMD results. The buffer size for $(17 \times 0)$ tube converges to desired accuracy much later than $(10 \times 10)$ tube. This results in much longer simulation time.

In the Figs. \ref{4.1.3} and \ref{4.1.4}, the effect of the buffer size 
together with the varying electronic temperatures (from $k_BT=0.000001~eV$ 
($\approx$0.012 K) to $k_BT=0.1~eV$ ($\approx$1200 K)) on the O(N) TBMD 
total energy value for the $(10 \times 10)$ and $(17 \times 0)$ tube 
structures are also given. It  is seen that when the buffer size value is 
small, electronic temperature has a slight effect on the energy value.
These values are the static results without performing simulation. The effect of the electronic temperature may be impressive during the simulation, when the forces between the atoms become dynamic. Therefore during simulations, it is safe to choose the electronic temperature as room temperature.

\begin{figure}
 \vskip 18.0 cm
    \includegraphics{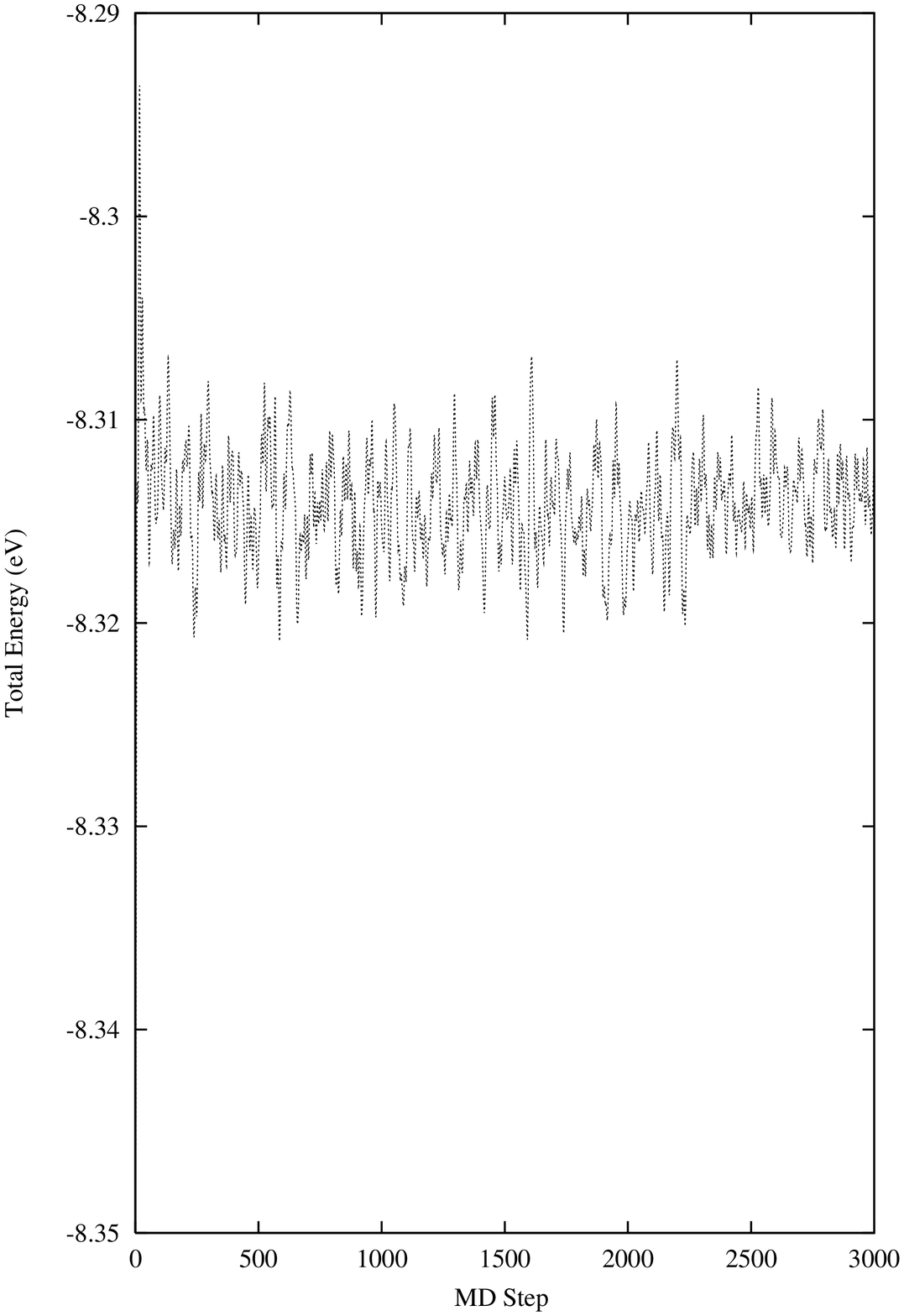}
\vskip -18.0 cm $\left. \right.$
\vspace{2.5cm}

\vskip 18.0 cm
    \includegraphics{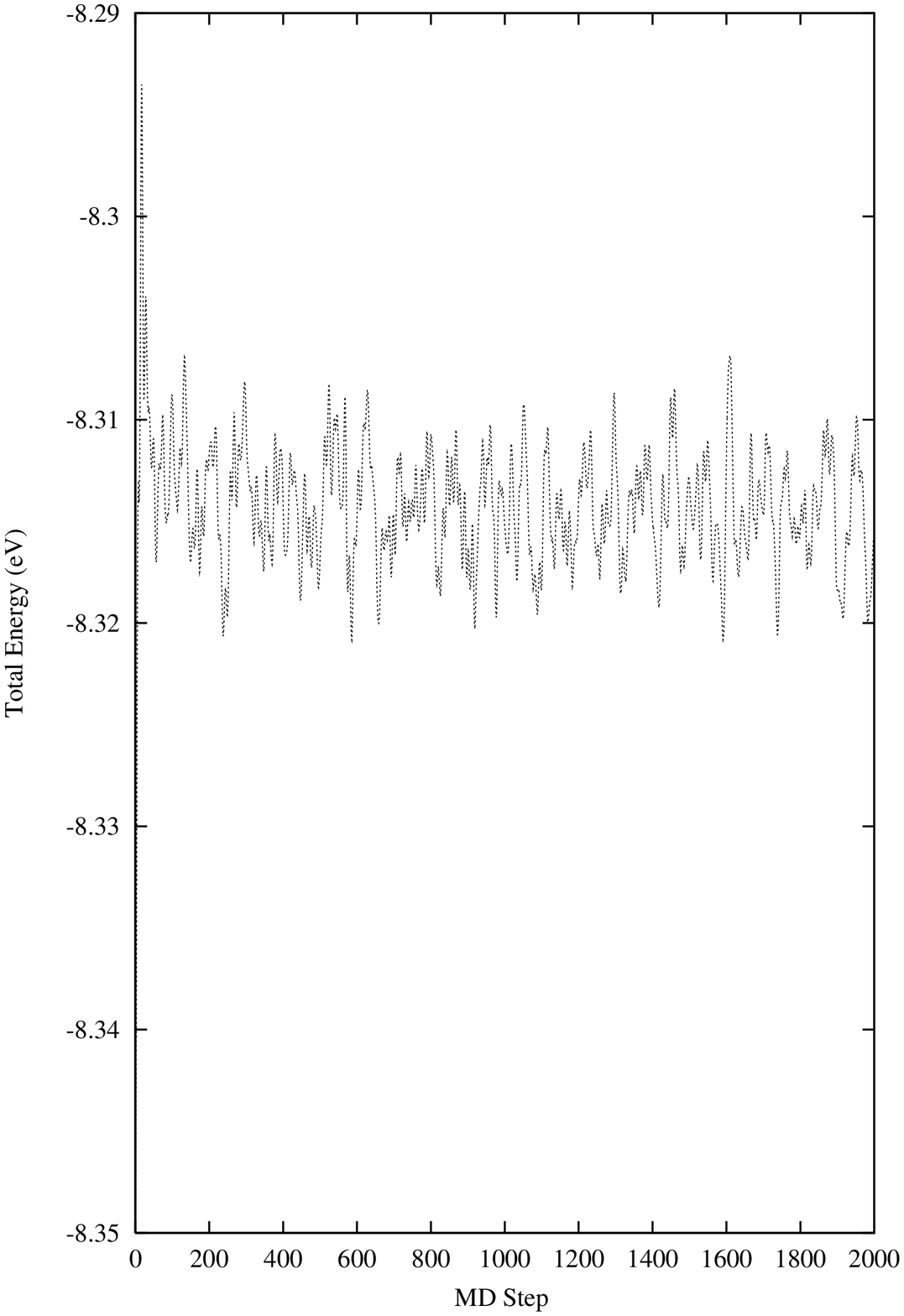}
\vskip 1 cm
\caption{The effect of electronic
temperature on the total energy  (Tube 10x10 no strain   T= 300  K)
for  the values  $k_BT=0.025~eV$ and
$k_BT=0.05~eV$, respectively.}
\vspace{2.5 cm}
\label{4.1.2}
\end{figure}

Next the effect of the electronic temperature on the total energy is investigated. 
The energetics for the pristine $(10 \times 10)$ tube with different electronic temperature 
values  $k_BT=0.025~eV$ ($\approx$300 K) and
$k_BT=0.05~eV$ ($\approx $600 K) are studied. Results are given 
in the Fig. \ref{4.1.2}. The upper graph is for the room temperature and the lower is the twice of the room temperature. Having an equal average energy value both graphs show similar behavior. They fluctuate around almost the same value. This is reasonable because they both simulate the same system with different electronic temperatures. The pattern at the above graph is more dense than
the lower one. This is because of the hotter electrons in the system.

\begin{figure}
 \vskip 18.0 cm
    \includegraphics{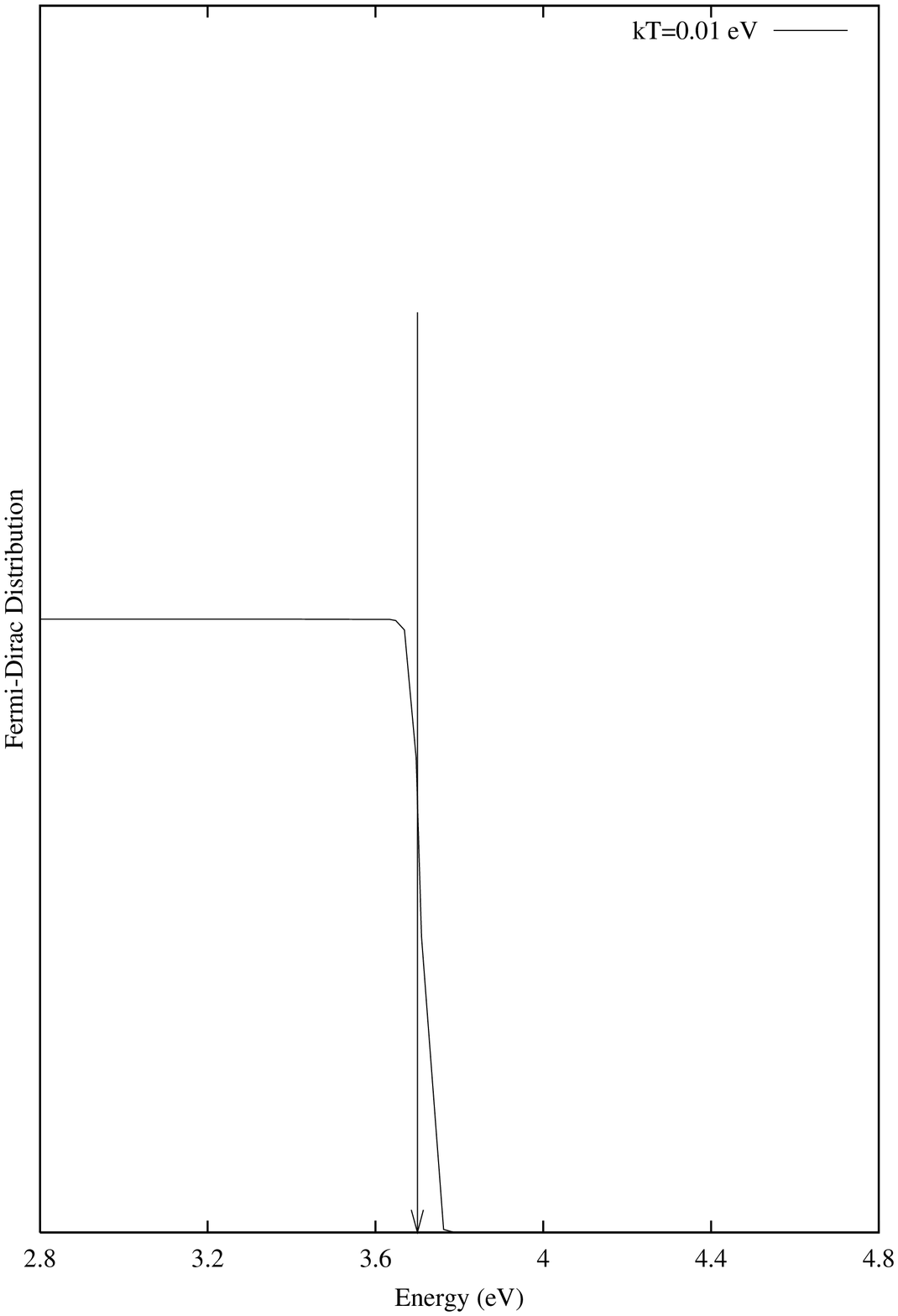}
\vskip -18.0 cm $\left. \right.$
\vspace{2.5cm}

\vskip 18.0 cm
    \includegraphics{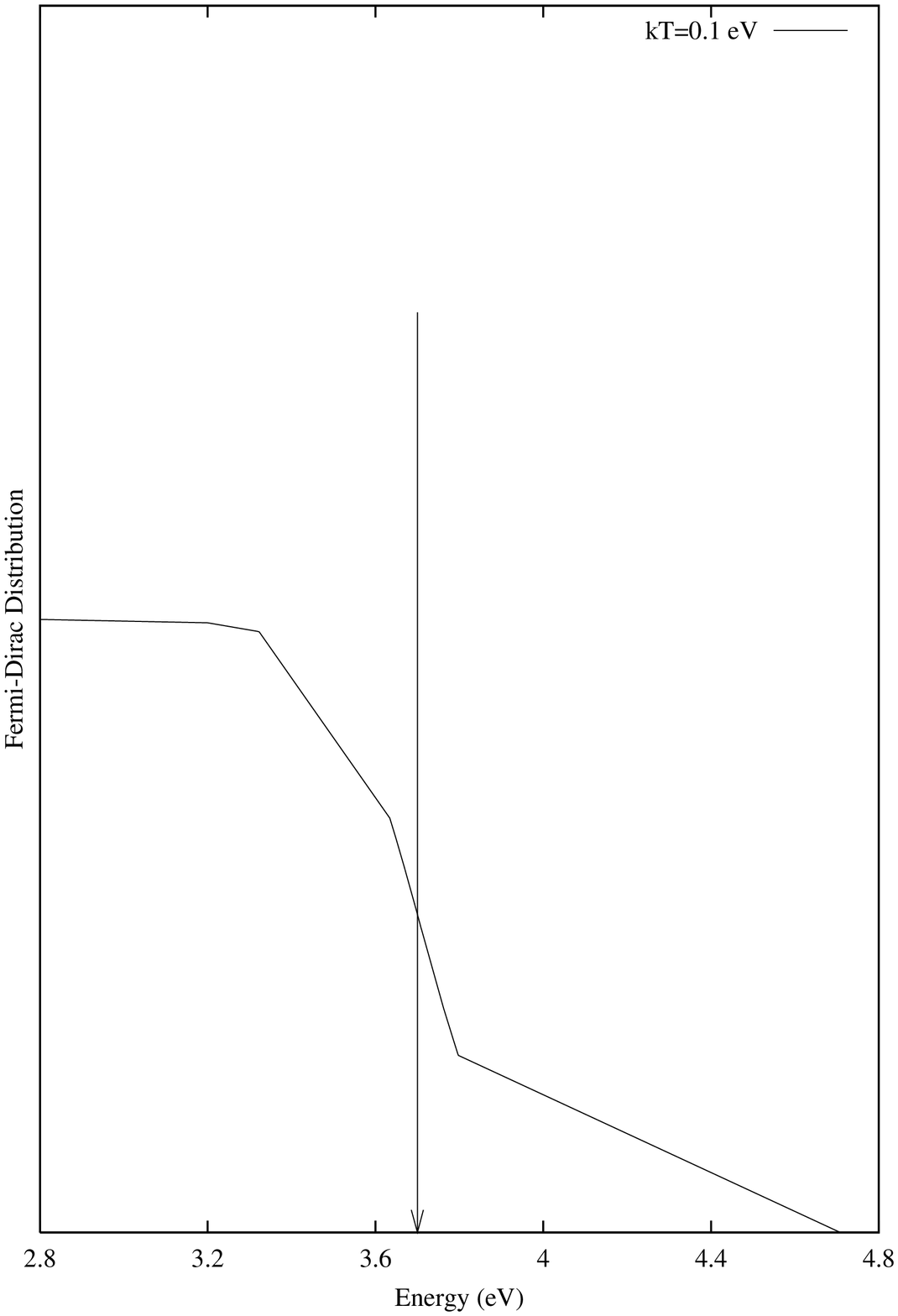}
\vskip 1 cm
\caption{Fermi-Dirac Distribution Function vs Energy for various
electronic temperatures  for the Tube Structure 10x10.}
\vspace{2.5 cm}
\label{4.1.5}
\end{figure}

\begin{figure}
 \vskip 18.0 cm
    \includegraphics{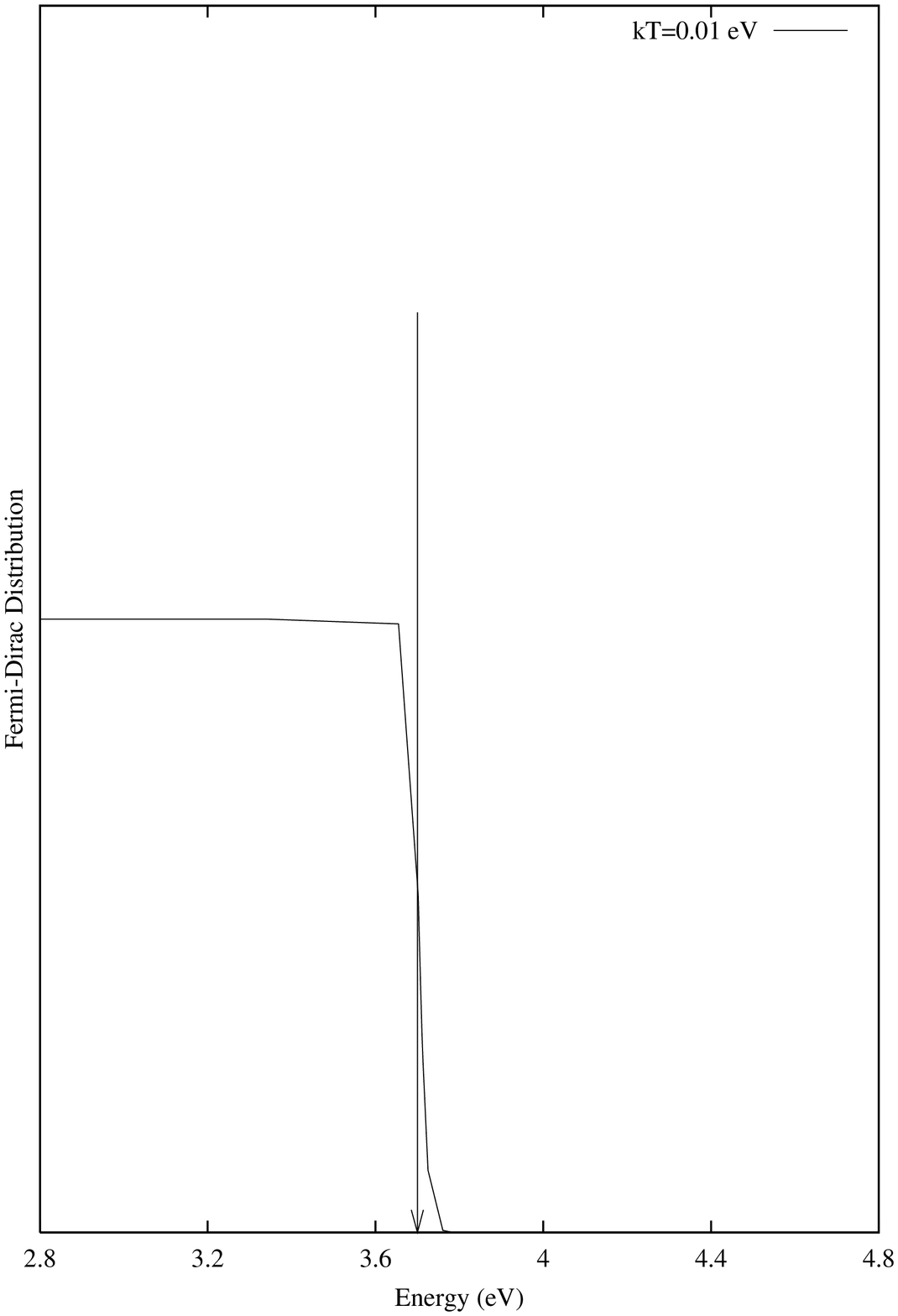}
\vskip -18.0 cm $\left. \right.$
\vspace{2.5cm}

\vskip 18.0 cm
    \includegraphics{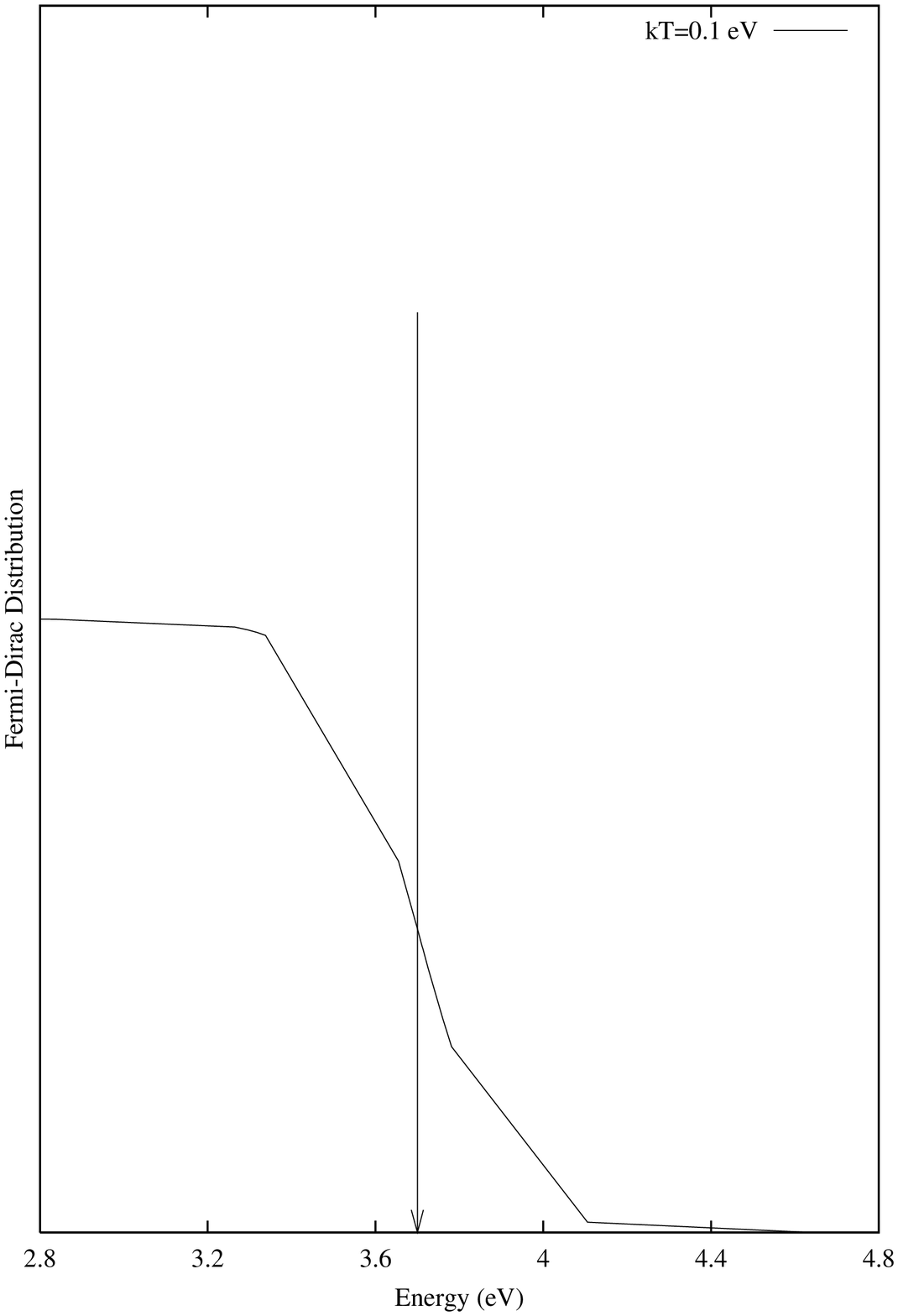}
\vskip 1 cm
\caption{Fermi-Dirac Distribution Function vs Energy for various
electronic temperatures  for the Tube Structure 17x0.}
\vspace{2.5 cm}
\label{4.1.6}
\end{figure}

In order to observe the effect of electronic temperature on the Fermi--Dirac Distribution of the pristine $(10 \times 10)$ and $(17 \times 0)$ tubes with different electronic temperature 
values  $k_BT=0.01~eV$ ($\approx$120 K) and
$k_BT=0.1~eV$ ($\approx $1200 K) are studied. Results are given 
in the Figs. \ref{4.1.5} and \ref{4.1.6}. The upper graphs in the figures are at 120 K while the lower ones  are at 1200 K. It is observed that as the electronic temperature is increased the graphs are broadening. 
Since less electronic state is populated at the low electronic temperature condition there is no widening for the upper graphs as expected.

\begin{figure}
 \vskip 18.0 cm
    \includegraphics{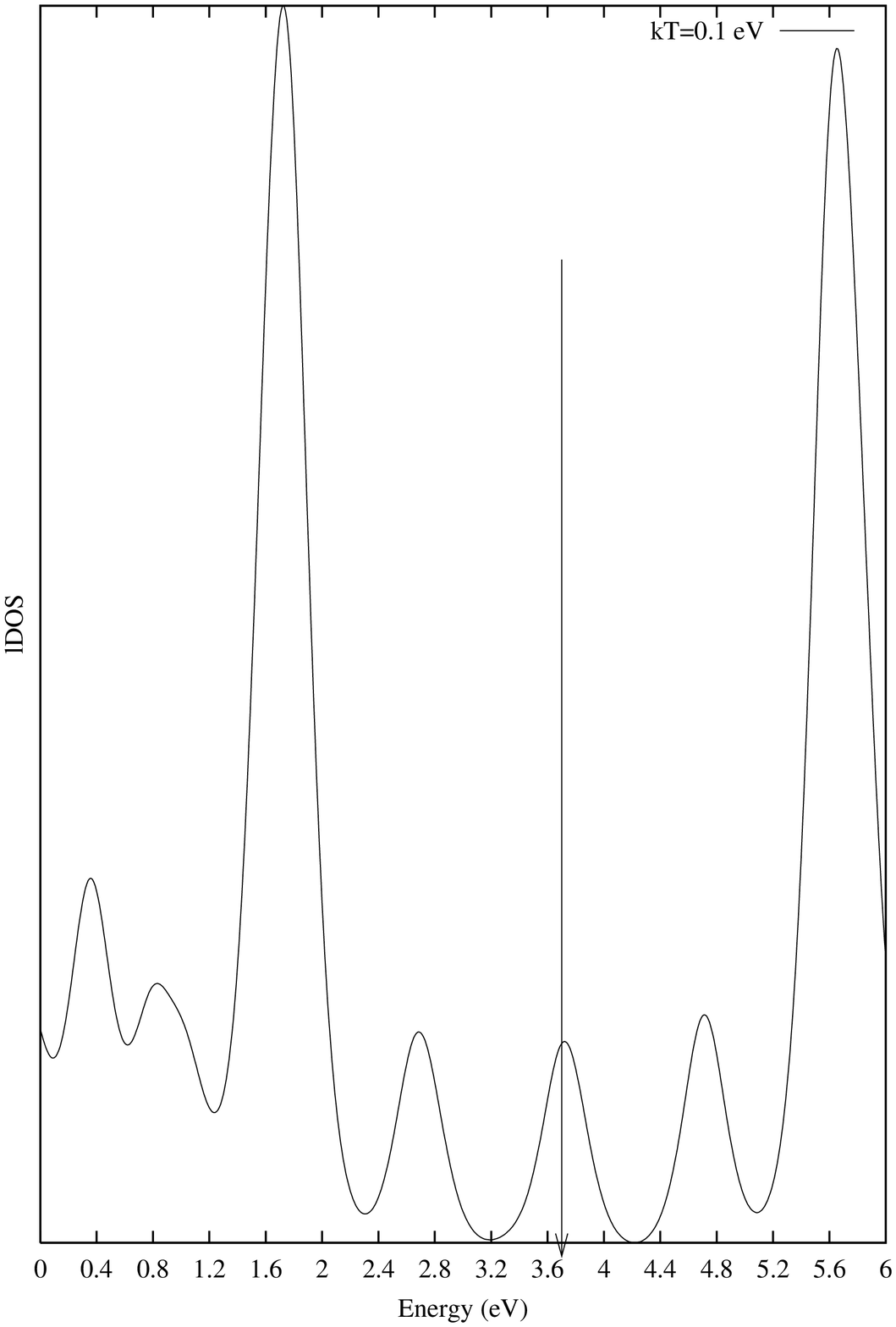}
\vskip -18.0 cm $\left. \right.$
\vspace{2.5cm}

\vskip 18.0 cm
    \includegraphics{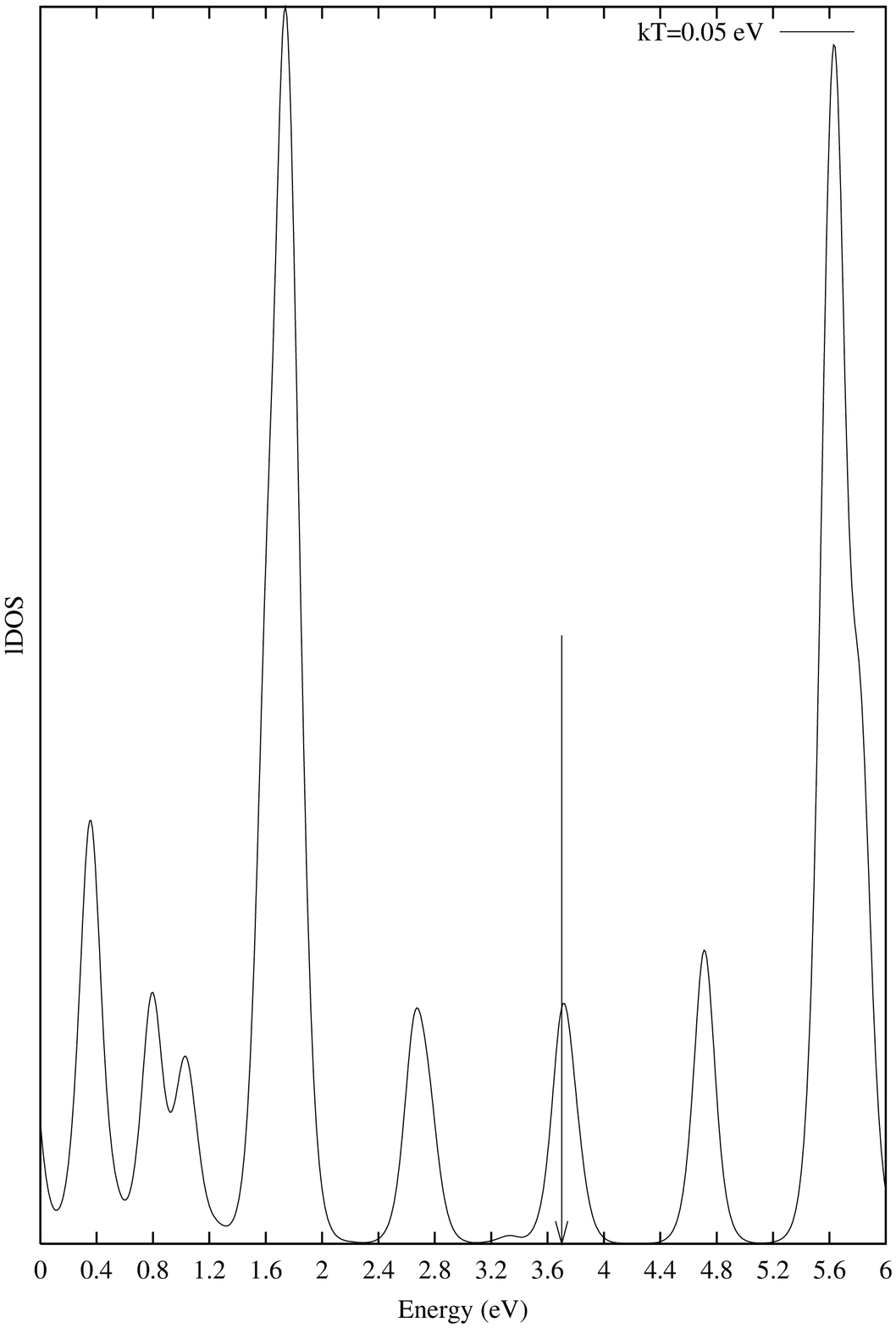}
\vskip 1 cm
\caption{Local DOS vs Energy for different electronic temperatures  for the Tube Structure 10x10..}
\vspace{2.5 cm}
\label{4.1.7}
\end{figure}

\begin{figure}
 \vskip 18.0 cm
    \includegraphics{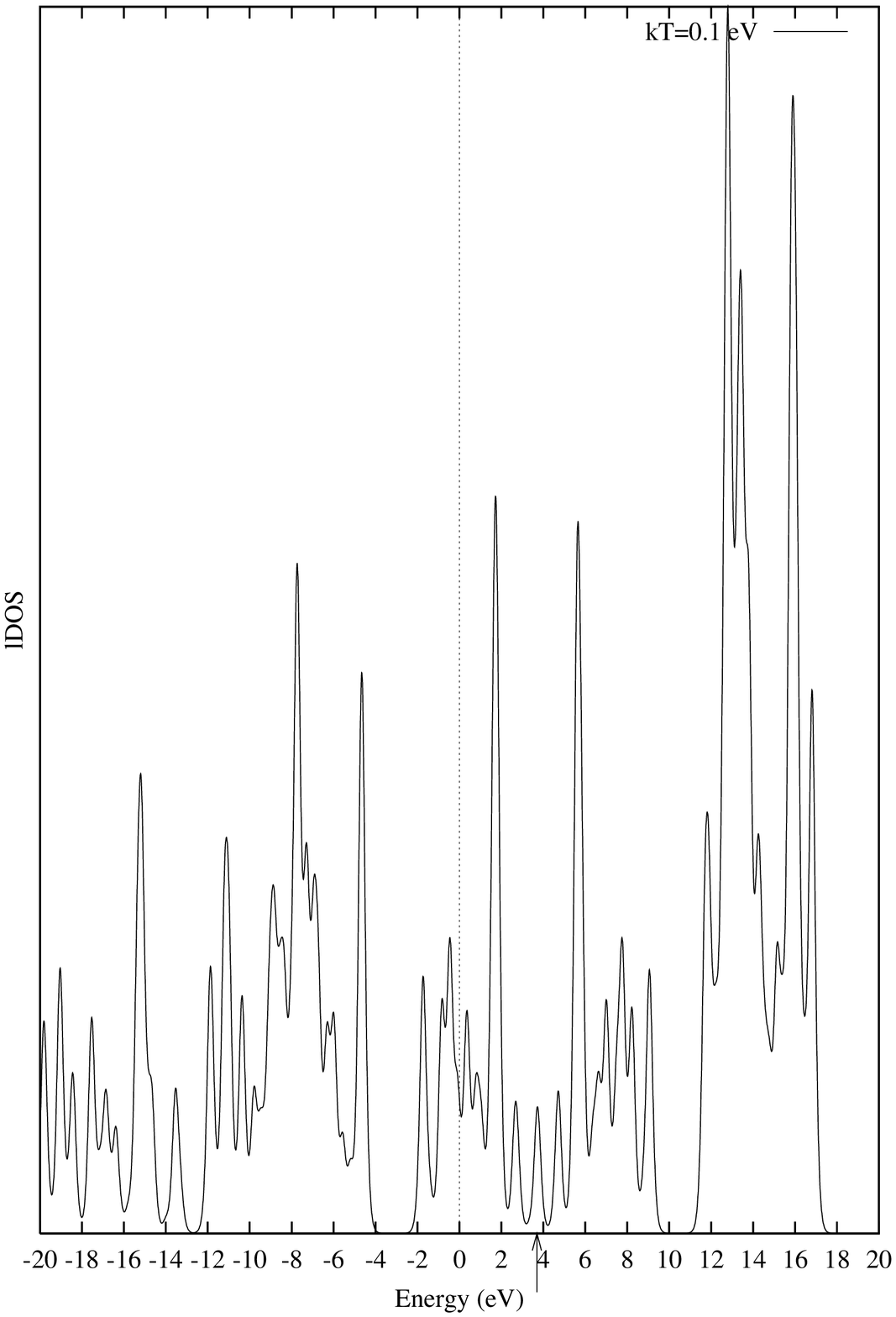}
\vskip -18.0 cm $\left. \right.$
\vspace{2.5cm}

\vskip 18.0 cm
    \includegraphics{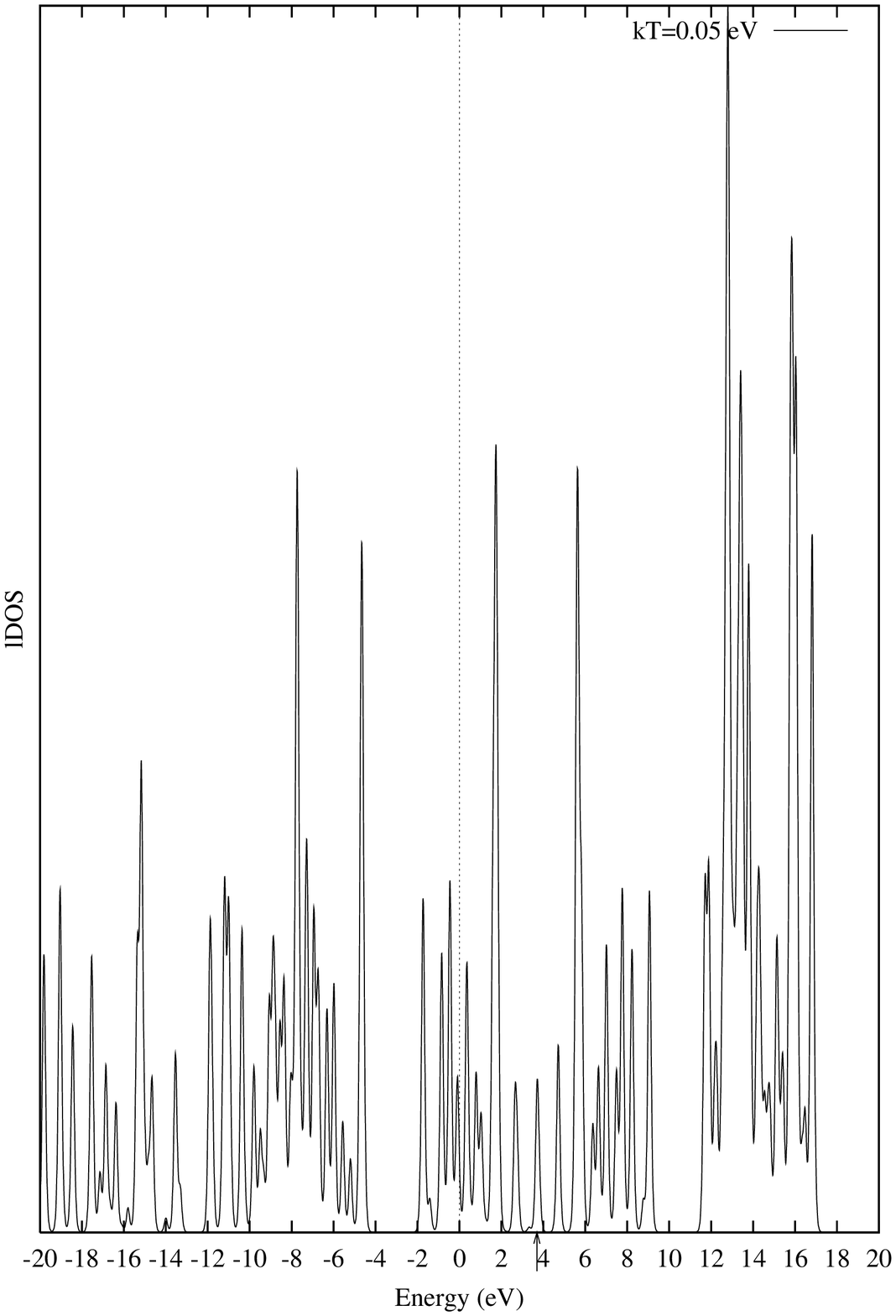}
\vskip 1 cm
\caption{Local DOS vs Energy for different electronic temperatures  for the Tube Structure 10x10}
\vspace{2.5 cm}
\label{4.1.8}
\end{figure}

\begin{figure}
 \vskip 18.0 cm
    \includegraphics{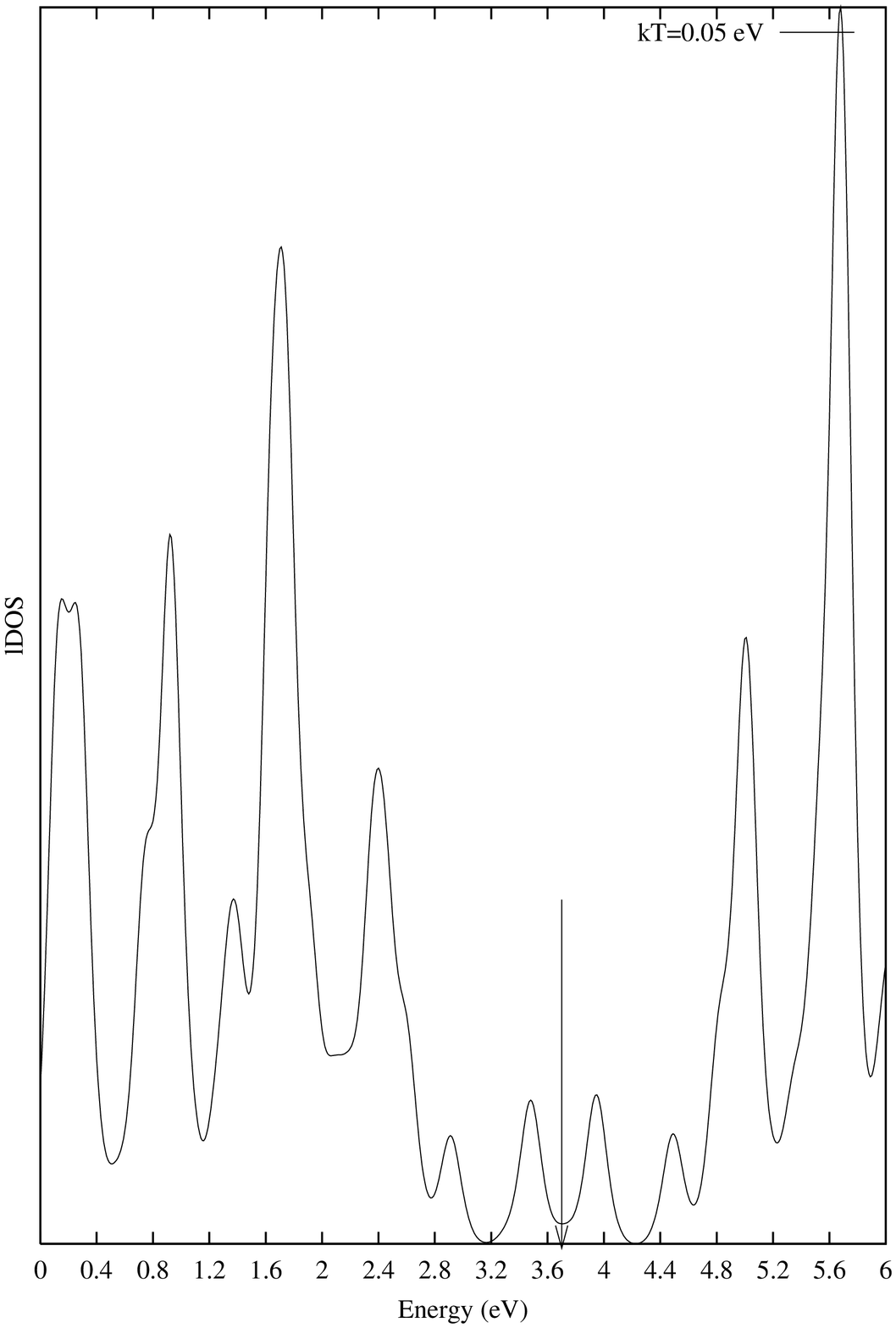}
\vskip -18.0 cm $\left. \right.$
\vspace{2.5cm}

\vskip 18.0 cm
    \includegraphics{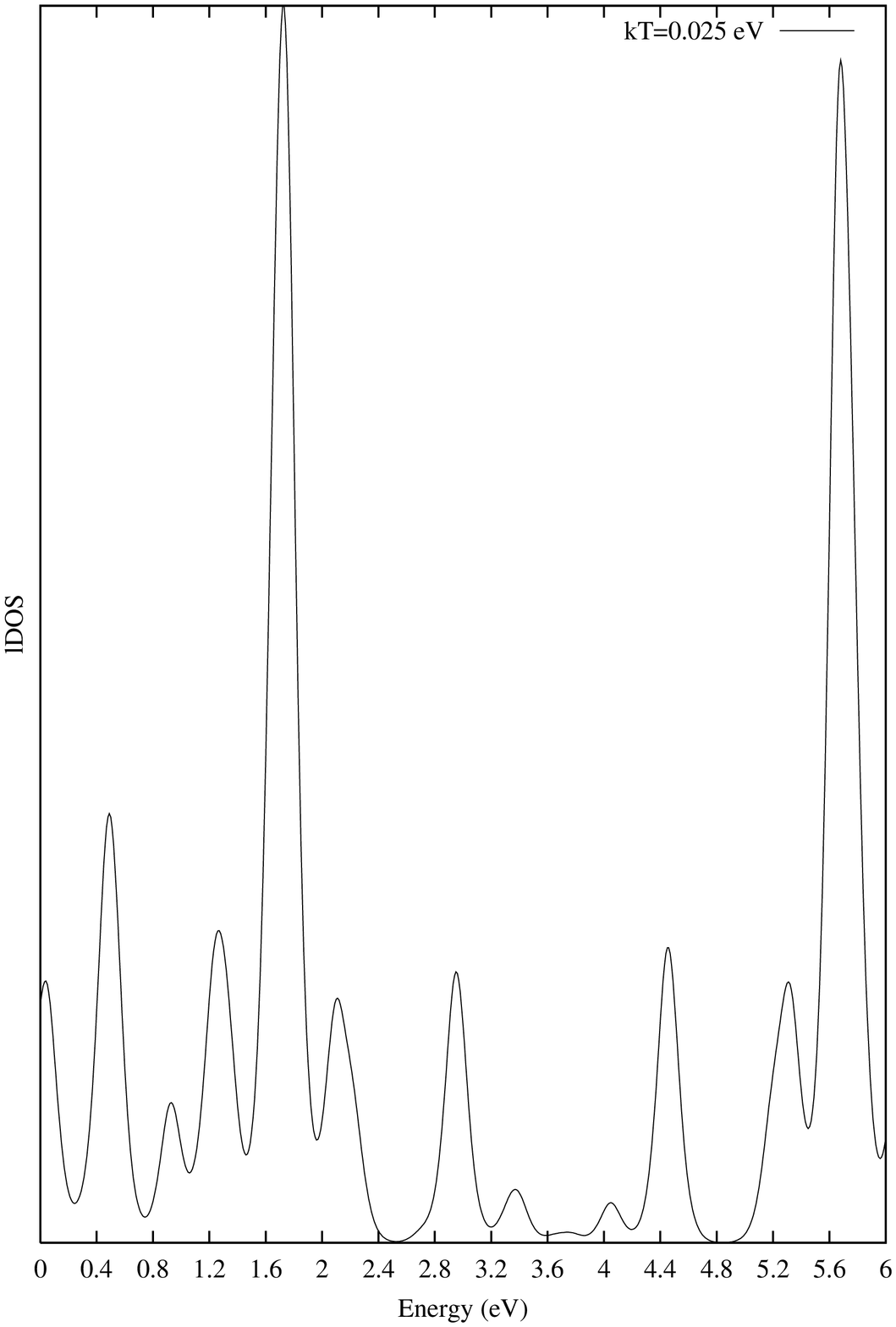}
\vskip 1 cm
\caption{
Local DOS vs Energy for different electronic
temperatures  for the Tube Structure 17x0.}
\vspace{2.5 cm}
\label{4.1.9}
\end{figure}

\begin{figure}
 \vskip 18.0 cm
    \includegraphics{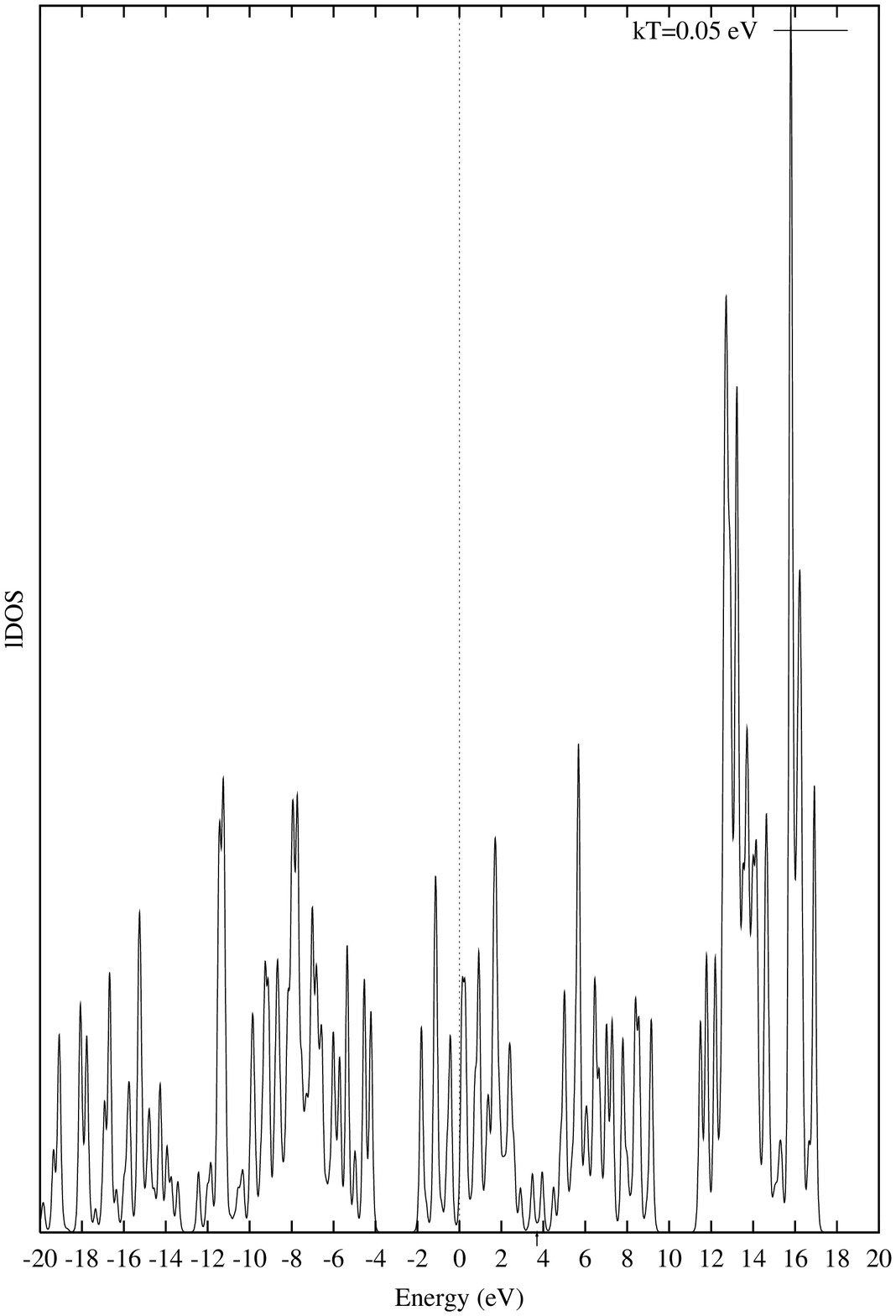}
\vskip -18.0 cm $\left. \right.$
\vspace{2.5cm}

\vskip 18.0 cm
    \includegraphics{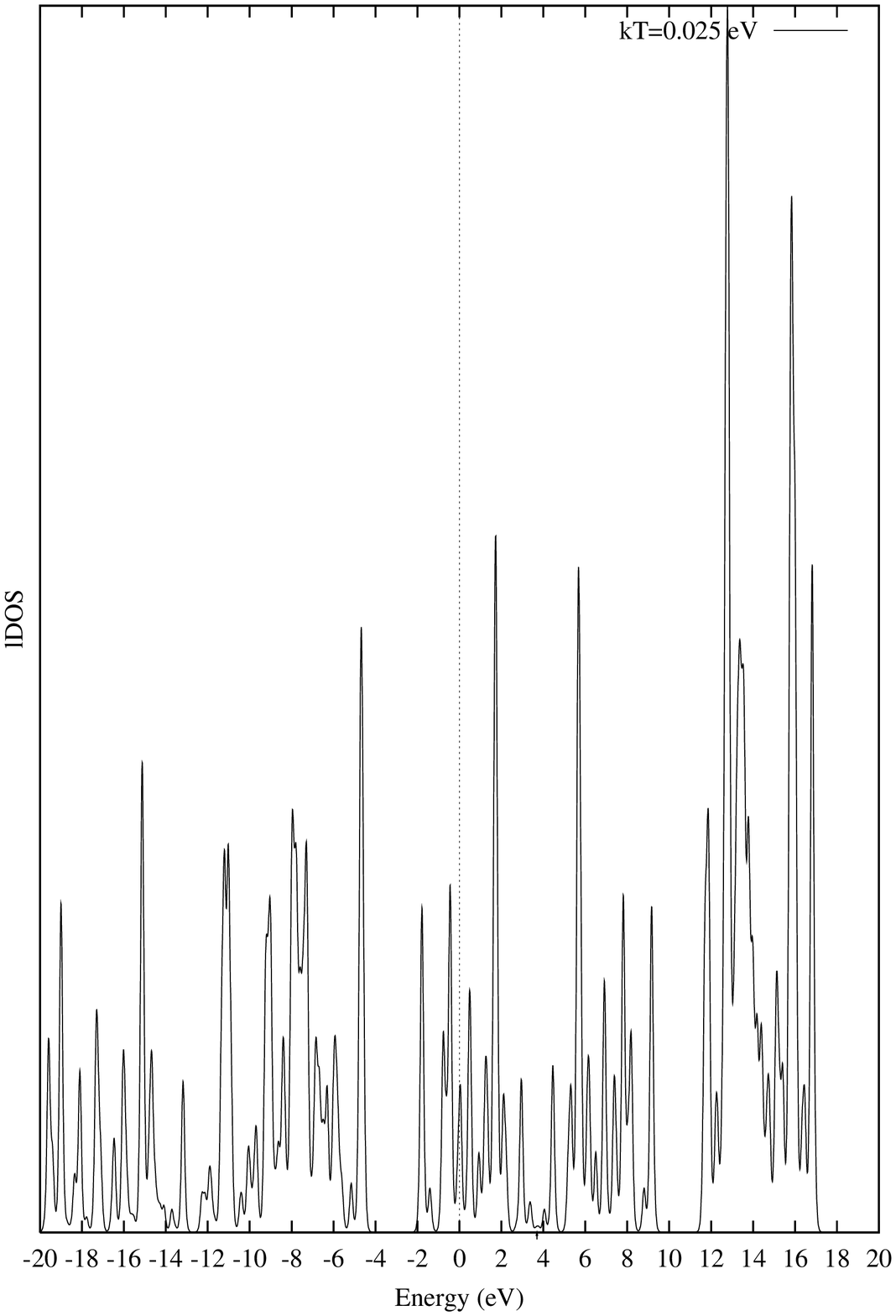}
\vskip 1 cm
\caption{
Local DOS vs Energy for different electronic
temperatures  for the Tube Structure 17x0.}
\vspace{2.5 cm}
\label{4.1.10}
\end{figure}

The density of states is obtained from the general formula,
\begin{equation}
g({\cal{E}\/})={dN({\cal{E}\/}) \over d{\cal{E}\/}}= {N({\cal{E}\/}+\epsilon)-N({\cal{E}\/})
 \over \epsilon}
\label{eq4.1.1}
\end{equation}
where N is the number of electrons in the system and equal to,
\begin{equation}
N(E)=\sum^{NCell} \sum^{4\cal{N}\/}_{i=1}
 {2 \over 1+f((E- {\cal E}_i)/k_BT)} *\sum_{j=1}^{4N} |H(j,i)]^2
\end{equation}
In the Eq. \ref{eq4.1.1}, the statement is that; if there is a change in the slope this gives us the information about the existence of populated electronic state. The criteria is the change in the slope for the Figs. \ref{4.1.7} -- \ref{4.1.10}.

In the Figs. \ref{4.1.7} -- \ref{4.1.10}, the lDOS graphs for the pristine $(10 \times 10)$ and $(17 \times 0)$ tubes for different electronic temperature values  $k_BT=0.1~eV$ ($\approx$1200 K), $k_BT=0.05~eV$ ($\approx $600 K) and  $k_BT=0.05~eV$ ($\approx$600 K),
$k_BT=0.025~eV$ ($\approx $300 K); respectively are given. In the Figs. \ref{4.1.7} and \ref{4.1.9}, only a selected range for lDOS are given to better understanding of the behavior of electronic states near the Fermi--Energy level for the tube structures $(10 \times 10)$ and $(17 \times 0)$, respectively. The other figures, namely \ref{4.1.8} and \ref{4.1.10}, give the same information but in the full range. It is seen that when the electronic temperature is increased the graphs begin to be smoother since higher amount of electronic states are populated. But, the peaks at and around the Fermi energy level are at the same positions for different electronic temperatures as expected. 

The Fermi energy level values are very similar (around 3.7 eV) for  both tube structures.  Although they have different chirality this is expected because two tubes have the same radii. The formula for the DOS gives the electronic state population for the  different energy values. It is found that the $(10 \times 10)$ tube has metallic behavior since it has states around Fermi energy level and a wide band gap but on the other hand the $(17 \times 0)$ tube has semiconducting
behavior since it has no states around Fermi energy level and small band gap as expected.

We have also calculated the band gap values for $(10 \times 10)$ and $(17 \times 0)$ tubes as 2.01 eV and 0.53 eV; respectively. In the literature \cite{wiveri}-\cite{ouchli}, the proposed model values are calculated by the formulas $2\gamma_0a_{c-c}/d$ and $6\gamma_0a_{c-c}/d$ (where $\gamma_0=2.5-2.7~eV$, $a_{c-c}=0.142~nm$, and d for diameter in (nm)) for semiconducting and metallic tubes, respectively. A theoretical value for $\gamma_0$ of 2.5 eV has been estimated by \cite{minwhi} using a first--principles local density approximation (LDA) to calculate the band structure of armchair carbon nanotube. As it is discussed in \cite{vemela} this type of calculations give 10--20\% smaller value for armchair nanotubes. The band gap values for the $(10 \times 10)$ and $(17 \times 0)$ Carbon Nanotubes using this model are 1.62--1.75 eV and 0.54--0.58; respectively. Our O(N) TBMD algorithm gives good energy band gap result for the $(17 \times 0)$ tube. For the $(10 \times 10)$ armchair tube the model value is lower than our calculated value.  
On the other hand, the behaviors of the local density of states graphs are (see Figs. \ref{4.1.7}--\ref{4.1.10}) as expected. For the 10x10 tube (metallic behavior), band gap  is wide and there are states populated around Fermi energy level and for the $(17 \times 0)$ tube (semiconducting behavior), band gap  is narrow and no states around Fermi energy level as expected.

\section{Conclusion}

In this study, the details of O(N) TBMD algorithm is given. It is described that how a system is divided into many subsystems and how their contributions give overall system properties (such as charge density, band structure energy) 
by using nearsightness principle. This principle uses the approximation that only the local environment contributes to the bonding of each atom. This gives us the opportunity for linear scaling. The main problem in the traditional TB is the increasing system size. When the system size increases (N), the time to diagonalize the constructed Hamiltonian matrix becomes in the order of $N^3$. The O(N) algorithms overcome this bottleneck and the behavior has a linear scaling. It is shown in Ref. \cite{ozdeca} that our O(N) algorithm scales linearly for increasing system size.

To conclude, our methodology is able to produce the physical properties such as Fermi-Dirac Distribution, local Density of  States and energetics for the Carbon Nanotubes. The structural stability under extreme conditions such as uniaxial strain
will be studied separately.

\begin{acknowledgments}
We thank Dr. Tahir \c{C}a\u{g}{\i}n (CALTECH) for discussions and his help
with $O(N)$ algorithms during his TOKTEN/UNISTAR visit.  The research reported here is supported by
T\"{U}B\.{I}TAK (The Scientific and Technical Research Council of Turkey) through the project TBAG-1877 
and by the Middle East Technical University through the project AFP-2000-07-02-11.
\end{acknowledgments}

\end{document}